\documentclass[%
reprint,
superscriptaddress,
groupedaddress,
nofootinbib,
amsmath,amssymb,
aps,
prd,
floatfix,
showkeys,
]{revtex4-2}

\usepackage{graphicx}
\usepackage{dcolumn}
\usepackage{bm}
\usepackage{hyperref}
\hypersetup{
    colorlinks=true,
    allcolors=blue,
}

\newcommand{\mycom}[2]{\left[#1,\, #2\right]}

\newcommand{\myeq}[1]{Eq.~\eqref{#1}}
\newcommand{\mysec}[1]{Sec.~\ref{#1}}
\def\ocal{\mathcal{O}}
\def\oprime{\ocal'}
\def\elsv{E_\text{LSV}}
\def\massbox{M_\text{box}}

\begin{document}


\title{On the Consistency of Covariant Light-Speed Variation in Doubly Special Relativity}

\author{Hao Li}
 \email{haolee@cqu.edu.cn}
\author{Jie Zhu}%
 \email{Corresponding author: jiezhu@cqu.edu.edu}
\affiliation{%
 School of Physics, Chongqing University, Chongqing 401331, P.R. China.
}%

\date{\today}


\begin{abstract}
Doubly special relativity (DSR) introduces an observer-independent energy scale while preserving a deformed relativistic notion of covariance. In many realizations, this leads to an energy-dependent speed of light (light-speed variation, LSV). We investigate the consistency of such observer-independent LSV through a thought experiment involving an inertial box emitting two photons in opposite directions.
We first distinguish two classes of LSV scenarios: those with the standard velocity-composition law, and those with observer-independent speed-energy relations, as in DSR. Focusing on the latter, we perform a quantitative analysis within the DSR1 model based on the $\kappa$-Poincar\'e algebra.
In the subluminal case ($\ell<0$), we derive a critical rapidity above which the boosted box overtakes its own photon, and show that this rapidity is physically attainable even after taking macroscopic effects into account. Within a standard particle interpretation, this leads to tensions in particle counting and inertial motion across frames. Unlike previously discussed issues in DSR, this effect does not appear to be resolvable by relative locality alone.
Our results point to a structural tension among observer-independent LSV, relativistic covariance, and standard notions of particle propagation in DSR frameworks.
\end{abstract}

                        
\maketitle


\section{Introduction\label{sec:intro}}

Special relativity~(SR) is built upon two central principles: the relativity principle and the universality of the speed of light.
While remarkably successful, SR does not incorporate any intrinsic energy or length scale.
This raises the question of how relativistic kinematics should be modified when a new fundamental scale---most naturally the Planck scale---is introduced.

Doubly special relativity~(DSR) has been proposed as a minimal extension of SR that preserves the relativity principle while introducing an additional observer-independent scale~\cite{Amelino-Camelia:2000stu,Amelino-Camelia:2000cpa,Kowalski-Glikman:2002iba,Magueijo:2001cr,Magueijo:2002am,Amelino-Camelia:2002uql,Kowalski-Glikman:2006ssl}.
In many realizations, this leads to a modified dispersion relation and, consequently, an energy-dependent photon propagation speed, often referred to as light-speed variation~(LSV).
Such effects have been widely studied in quantum-gravity phenomenology, for instance, through time-of-flight analyses of high-energy astrophysical signals~\cite{Mattingly:2005re,Amelino-Camelia:2008aez,Addazi:2021xuf,AlvesBatista:2023wqm}.

Despite these phenomenological applications~(see, e.g.,~\cite{FermiGBMLAT:2009nfe,Amelino-Camelia:2008aez,Addazi:2021xuf,AlvesBatista:2023wqm}), the internal consistency of observer-independent LSV remains subtle.
If the functional dependence of the speed of light on energy is required to be the same in all inertial frames, the interplay between deformed boost transformations and particle kinematics may lead to nontrivial conceptual tensions.
Previous works, notably by Hossenfelder~\cite{Hossenfelder:2009mu,Hossenfelder:2010tm}, have identified potential inconsistencies in such scenarios.
Notably, these paradoxes can be understood, at least in part, as manifestations of relative locality~\cite{Amelino-Camelia:2010wxx,Amelino-Camelia:2011lvm,Amelino-Camelia:2011ebd,Amelino-Camelia:2011dwc,Amelino-Camelia:2011hjg}.

In this work, we identify a distinct tension that does not appear to be resolved by appealing to relative locality alone.
We consider a simple but robust thought experiment: an inertial box emits two photons in opposite directions with equal and opposite momenta, such that the box remains inertial in its rest frame.
In standard SR, the forward photon always outruns the box in any boosted frame.
However, with observer-independent LSV, sufficiently large boosts can reverse this ordering, so that the box overtakes its own emitted photon in a boosted frame.
The core of the effect is thus a reversal of the ordering between signal propagation and source motion under boosts.

Within a standard particle interpretation, this leads to tensions that cannot be straightforwardly dismissed as coordinate artifacts.
Different inertial observers may disagree on the number of propagating particles and on whether the box remains inertial.
We adopt the operational definition that particle number is determined by asymptotic detection and must therefore be invariant under frame transformations.
Under this assumption, the above scenario signals a tension with standard expectations of relativistic consistency.

To clarify the origin of this tension, we distinguish two broad classes of LSV scenarios.
In the first class, the velocity composition law remains that of SR, and any energy dependence of the speed is confined to a preferred frame (sometimes interpreted as an \ae{}ther frame).
In this case, no tension of the type discussed here arises.
In the second class, the functional form of the speed-energy relation is observer-independent, with covariance maintained through deformed transformations.
This is the defining feature of DSR, and it is in this class that the tension becomes manifest.

We then perform a detailed quantitative analysis within the well-studied DSR1 model based on the $\kappa$-Poincar{\'e} algebra.
In the subluminal case, we derive a critical rapidity above which the box overtakes its own emitted photon, and demonstrate that this rapidity lies within the physically allowed range of deformed boosts, even when macroscopic effects are taken into account.
In the superluminal case, similar effects can arise but may be avoided under additional assumptions.

Our results point to a structural tension among three ingredients: observer-independent LSV, relativistic covariance, and standard notions of particle propagation.
Crucially, unlike the paradoxes identified in Refs.~\cite{Hossenfelder:2009mu,Hossenfelder:2010tm}, the effect we identify persists even when relative locality is taken into account.
We show in detail that relative locality does not, by itself, resolve the tension and instead introduces additional ambiguities in particle interpretation and causality.

The remainder of this paper is organized as follows.
In \mysec{sec:model-indep} we present the model-independent setup and classify different LSV scenarios.
In \mysec{sec:intro-dsr} we briefly review the DSR1 framework.
\mysec{sec:analysis} contains the detailed analysis, and \mysec{sec:discussion} concludes with a discussion of implications and possible extensions.


\section{Model-Independent Analysis \label{sec:model-indep}}

\begin{figure}
    \centering
    \includegraphics[width=\linewidth]{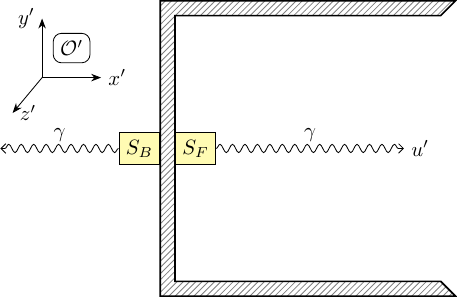}
    \caption{Schematic diagram of the box~(laboratory) at rest in the frame \(\oprime\).
    At some time, the source \(S_F\)~(\(S_B\)) emits a photon in the \(+x\)-direction~(\(-x\)-direction), while the two sources are adjusted such that the box always remains at rest.}
    \label{fig:fig1}
\end{figure}

\begin{figure}
    \centering
    \includegraphics[width=\linewidth]{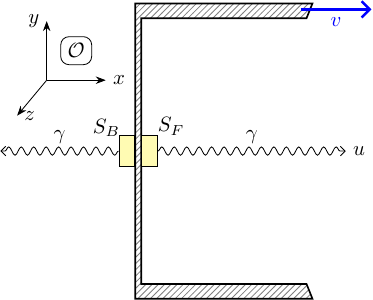}
    \caption{Schematic diagram of the box in Fig.~\ref{fig:fig1}, but viewed in the frame \(\ocal\), where the box moves in the \(+x\)-direction with velocity \(v\). This figure assumes the ordinary rule of special relativity; thus, from the perspective of an observer at rest in \(\ocal\), both sources emit photons in opposite directions.}
    \label{fig:fig2}
\end{figure}

We begin by reviewing the relevant aspects of special relativity~(SR).
We construct a thought experiment as follows.
Consider a box (for example, a laboratory, though the details are irrelevant) located at the origin of an inertial frame \(\oprime\). Two photon emitters (sources) \(S_F\) and \(S_B\) are placed on opposite sides of the left wall, as depicted schematically in Fig.~\ref{fig:fig1}.
The precise emission mechanism is unimportant; one may take the process \(\pi^0\to\gamma\gamma\) as a concrete example.
Without loss of generality, at \(t'=0\), let \(S_F\) emit a photon forward along the \(+x'\) axis, and simultaneously let \(S_B\) emit a photon backward such that the box remains at rest.
In SR the two photons must have the same energy, although the backward-propagating photon plays no role in our analysis.
For observers in the box frame \(\oprime\), the forward photon propagates to infinity and the box can never catch up with it.
Now consider another inertial frame \(\ocal\), in which the box moves in the \(+x\)-direction with speed \(v\), as shown in Fig.~\ref{fig:fig2}.
It is a well-known fact that, no matter how fast the box moves in this frame, it can never catch up with the forward-moving photon.
Indeed, the speed \(u\) of the forward photon in this frame satisfies
\begin{equation}
    u=\frac{v+ u'}{1+vu'}=1>v,\label{eq:velocitycomp}
\end{equation}
where \(u'=1\) is the speed of light in the frame \(\oprime\).
This is simply the principle of special relativity: the speed of light is the same constant in every inertial frame, and it is the maximum attainable speed.

\begin{figure}
    \centering
    \includegraphics[width=\linewidth]{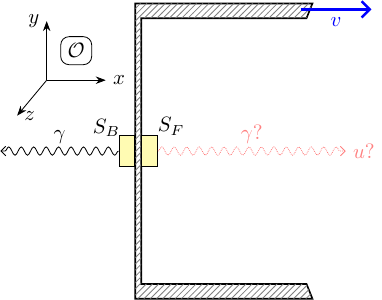}
    \caption{Schematic diagram of the box as in Fig.~\ref{fig:fig2}. In this diagram the effects of light-speed variation are taken into account, and from the viewpoint of an observer in \(\ocal\), it is questionable whether the source \(S_F\) ever emitted a forward-moving photon.}
    \label{fig:fig3}
\end{figure}

\begin{figure}
    \centering
    \includegraphics[width=\linewidth]{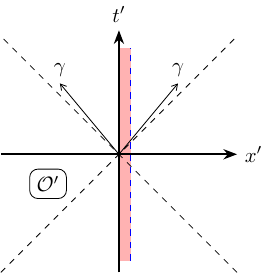}
    \caption{Spacetime diagram of Fig.~\ref{fig:fig1}, where the filled region represents the worldline of the box, and the worldlines of the two photons have slopes of equal magnitude.}
    \label{fig:fig4}
\end{figure}

\begin{figure}
    \centering
    \includegraphics[width=\linewidth]{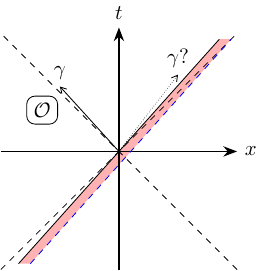}
    \caption{Spacetime diagram in \(\ocal\) corresponding to Fig.~\ref{fig:fig2} or Fig.~\ref{fig:fig3}. Because of light-speed variation, one must carefully analyze whether the forward-moving photon is faster or slower than the box in \(\ocal\).}
    \label{fig:fig5}
\end{figure}

In scenarios with light-speed variation~(LSV), however, the situation can be very different, and a tension may arise.
Specifically, the problem emerges when \emph{in all inertial frames} the speed \(u\) of a photon varies linearly with its energy \(E\): \(u(E)\simeq 1-E/\elsv\), where \(E\ll \lvert\elsv\rvert\) and \(\elsv\) is the LSV scale, typically expected to be of the order of the Planck scale \(E_{Pl}\simeq 1.2\times 10^{19}~\text{GeV}\).
The subtlety is as follows.
Consider again the thought experiment described above, now taking LSV into account.
In the frame \(\oprime\), nothing prevents the forward-moving photon with speed \(u'\) and energy \(E'\) from moving away from the emitter.
Thus, for an observer at rest in \(\oprime\), there are always two photons once they are created and emitted.
But what does an observer at rest in \(\ocal\) see, as depicted in Fig.~\ref{fig:fig3}?
For a sufficiently boosted box (the frame \(\oprime\)) with speed \(v\) very close to unity, the energy \(E\) of the forward photon in \(\ocal\) becomes large.
Since its speed is \(u\simeq 1-E/\elsv\), it is possible that \(u<v\) if \(\elsv>0\).
If this occurs, the argument suggests that, within this description, the forward-moving photon may no longer correspond to a propagating signal in \(\ocal\), whereas the backward-moving photon is unaffected (see Figs.~\ref{fig:fig4} and~\ref{fig:fig5}).
An observer in \(\ocal\) would then conclude that only one photon contributes to backward asymptotic detection.

This thought experiment exposes a tension.
First, if the observer in \(\oprime\) and the observer in \(\ocal\) eventually meet and compare their conclusions about how many particles were created, they may arrive at different descriptions.
But particle number---defined operationally through asymptotic detection---is expected to be invariant under changes of inertial frame within any relativistic framework admitting a standard particle interpretation.
This discrepancy is therefore nontrivial.
Second, in \(\oprime\) the two photons are emitted such that the box remains at rest, whereas from the perspective of \(\ocal\) only one photon contributes to backward asymptotic detection, and the box appears to undergo an instantaneous acceleration.
The two observers thus also arrive at different descriptions of whether the box behaves as an inertial object.
Similar conclusions hold if the leading LSV correction is of higher order~\cite{Jacobson:2002hd}, \emph{e.g.,} \(u(E)\simeq 1-E^2/\elsv^2\).

The above arguments reveal the possibility of a tension inherent in subluminal LSV scenarios.
However, we must sharpen the logic, as there are conceptual ambiguities regarding how the speed of light depends on energy in different inertial frames.
We have assumed that in \(\ocal\) the speed of light also takes the form \(u(E)\simeq 1-E/\elsv\), implying that the functional form is observer-independent.
This need not be true, and we must carefully distinguish between the following two cases.
The first possibility is that the speed of light in different inertial frames is still related by \myeq{eq:velocitycomp}, in which case the tension never appears.
One can easily check that \(u-v=(1-v^2)u'/(1+vu')>0\), from which it follows that \emph{even if} photons are \emph{superluminal} in some frame, the inconsistency does not arise.
This case can be realized in models with an {\bfseries{\ae}ther}~(see, e.g.,~\cite{Coleman:1998ti,Mattingly:2005re}): there exists a special frame in which the speed of light varies with energy, while in other inertial frames speeds are composed according to the usual SR rule.
One may regard this special frame as the comoving frame of the \ae{}ther.
In this sense, the \ae{}ther provides a preferred frame in which the physical laws take a simple form.
The second possibility is that the \emph{functional form} of the speed of light is \emph{observer-independent}, while the composition law is deformed accordingly.
The most well-known realization of this case is doubly special relativity~(DSR), in which the algebraic structure of Lorentz symmetry is preserved, but the representation carried by the physical energy-momentum space (and by spacetime itself) is no longer linear.
If in \(\oprime\) and \(\ocal\) the observed energies of a photon are \(E'\) and \(E\), respectively, then the speeds in the two frames should be \(u'\simeq 1-E'/\elsv\) and \(u\simeq 1-E/\elsv\), respectively, with the LSV scale \(\elsv\) invariant.
If we boost the box sufficiently, it may effectively travel faster than the photon signal in certain frames within this description.
It is this second case---the DSR framework---on which we focus in this work.
The detailed analysis begins in the next section.
For clarity, we classify LSV scenarios as follows:
\begin{enumerate}
    \item[Case I] Among different inertial frames, the speed of light transforms according to \myeq{eq:velocitycomp}, while in a specific frame (the \ae{}ther frame) the speed has a prescribed functional dependence on energy.
    \item[Case II] No preferred frame exists, and physical laws in different frames are related by deformed ``covariant'' transformations. In this work we focus on doubly special relativity.
\end{enumerate}

\section{A brief introduction to DSR \label{sec:intro-dsr}}

In this section, we provide a brief introduction to doubly special relativity~(DSR).
Roughly speaking, DSR is a framework proposed to minimally extend special relativity~(SR) by introducing an additional invariant length scale \(\ell\)~\footnote{Equivalently, this length scale can be translated into an invariant energy scale, denoted by \(\elsv\), which represents the scale of new physics.}, which plays a role analogous to that of the speed of light in SR.
Thus, DSR is characterized by two invariants: the speed of light in vacuum (for photons with vanishing energy) and a minimal length \(\ell\).

We focus on one of the most extensively studied DSR models, DSR1~\cite{Amelino-Camelia:2000cpa,Amelino-Camelia:2000stu,Amelino-Camelia:2025ask}, which is inspired by the structure of the \(\kappa\)-Poincar{\'e} Hopf algebra~\cite{Lukierski:1991pn,Majid:1994cy,Kowalski-Glikman:2002iba}.
In DSR1, the usual Lie algebra of the Poincar{\'e} group is modified as follows (in the sense of Poisson brackets):
\begin{align}
    & \mycom{M_i}{M_j}=\epsilon_{ijk}M_k,\qquad \mycom{M_i}{N_j}=\epsilon_{ijk}N_k,\nonumber\\
    & \mycom{N_i}{N_j}=-\epsilon_{ijk}M_k,\label{eq:lorentz}
\end{align}
and
\begin{align}
    & \mycom{N_i}{P_j}=\delta_{ij}\left(\frac{1-e^{-2P_0\ell}}{2\ell}+\frac{\ell}{2}\boldsymbol{P}^2\right)-\ell P_iP_j,\nonumber\\
    & \mycom{P_\mu}{P_\nu}=0,\qquad \mycom{P_0}{N_i}=P_i,\label{eq:poincare}
\end{align}
where \(P_\mu\), \(M_i\), and \(N_i\) are the generators of translations, rotations, and Lorentz boosts, respectively.
One of the Casimirs of this algebra is
\begin{equation}
    C=\frac{4}{\ell^2}\sinh^2\left(\frac{\ell}{2}P_0\right)-\boldsymbol{P}^2e^{\ell P_0}.\label{eq:casimir1}
\end{equation}
Without loss of generality, in a \(1+1\)-dimensional phase space \((x^\mu,\, p_\mu)\), the Casimir yields the deformed \emph{mass-shell} condition:
\begin{equation}
    \frac{4}{\ell^2}\sinh^2\left(\frac{\ell}{2}p_0\right)-p_1^2 e^{\ell p_0}=m^2,\label{eq:mass-shell}
\end{equation}
with the rest mass (defined as the energy at zero momentum) given by \(M=2\sinh^{-1}(\ell m/2)/\ell\).

A finite boost transformation with rapidity \(\eta\) of the phase-space variables is given by~\cite{Bruno:2001mw,Amelino-Camelia:2025ask}
\begin{equation}
    \begin{aligned}
        p_0(\eta)&=\bar{p}_0\nonumber\\ 
        &\,-\frac{1}{\ell}\ln\frac{4e^\eta}{(1+e^\eta+\ell\bar{p}_1(e^\eta-1))^2-(e^\eta-1)^2e^{-2\ell\bar{p}_0}},\\
        p_1(\eta)&=\frac{2e^\eta(\ell^{-1}(\ell^2\bar{p}_1^2-e^{-2\ell\bar{p}_0}+1)\sinh\eta+2\bar{p}_1\cosh\eta)}{(1+e^\eta+\ell\bar{p}_1(e^\eta-1))^2-(e^\eta-1)^2e^{-2\ell\bar{p}_0}},\label{eq:momentrans}
    \end{aligned}
\end{equation}
for momenta, and~\cite{Amelino-Camelia:2025ask}
\begin{align}
    x^0(\eta)&= \bar{x}^0\frac{(1+e^\eta+\ell\bar{p}_1(e^\eta-1))^2+(e^\eta-1)^2e^{-2\ell\bar{p}_0}}{(1+e^\eta+\ell\bar{p}_1(e^\eta-1))^2-(e^\eta-1)^2e^{-2\ell\bar{p}_0}}\nonumber\\
    &\, - \bar{x}^1\frac{2e^{-2\ell\bar{p}_0}(1+e^\eta+\ell\bar{p}_1(e^\eta-1))(e^\eta-1)}{(1+e^\eta+\ell\bar{p}_1(e^\eta-1))^2-(e^\eta-1)^2e^{-2\ell\bar{p}_0}},  \nonumber\\
   x^1(\eta)&=\frac{\bar{x}^1}{4}e^{-\eta}\left[(1-\bar{p}_1\ell+e^\eta\bar{p}_1\ell+e^\eta)^2+(e^\eta-1)^2\right.\nonumber\\ 
   &\, \left.\times e^{-2\ell\bar{p}_0}\right]-\bar{x}^0(\sinh\eta+\bar{p}_1\ell\cosh\eta-\bar{p}_1\ell).\label{eq:coordtrans}
\end{align}
for spacetime coordinates.
Quantities with a bar denote the initial values at \(\eta=0\), i.e., the values in the unboosted frame.
Note that when \(\ell<0\), the rapidity cannot be arbitrarily large~\cite{Bruno:2001mw,Amelino-Camelia:2025ask}, and we will return to this issue later.


We are primarily interested in the speed of light in DSR1.
Although the definition of speed in DSR can be subtle, in this model, it can be given in a consistent and operationally meaningful way.
One can obtain the speed from \myeq{eq:mass-shell} by differentiating with respect to \(p_1\) and solving for \(\partial p_0/\partial p_1\).
Alternatively, one can define the coordinate speed as \(dx/dt\)~\cite{Daszkiewicz:2003yr,Amelino-Camelia:2011ebd,Mignemi:2019yzn,Mignemi:2003ab}.
These two definitions agree in this framework:
\begin{equation}
    V(\epsilon,p)=\frac{2e^{2\ell\epsilon}\ell p}{e^{2\ell\epsilon}(\ell^2p^2-1)+1}.\label{eq:speed}
\end{equation}
To leading order in \(\ell\), \myeq{eq:speed} reduces to the SR expression \(V(\epsilon,p)\simeq -p/\epsilon\)~\footnote{The leading behavior is \(V(\epsilon,p)\simeq -p/\epsilon+\mathcal{O}(\ell)\), which appears to differ by an extra sign from the usual result.
This is because we are using the covariant momentum \(p_\mu\); one must include an additional minus sign in either \(\epsilon\) or \(p\) to recover the correct physical interpretation (see also~\cite{Mignemi:2019yzn}).
Specifically, in this work a positive velocity (in the \(+x\)-direction) corresponds to negative \(p\) and positive \(\epsilon\).}.
Moreover, the kinematic velocity in DSR coincides with the usual definition of particle velocity, \(V_{\text{particle}}=\partial \epsilon/\partial p\), so within this framework there is no ambiguity in the definition of speed~\cite{Amelino-Camelia:2025ask}.

For massive objects, it is also useful to express the speed as a function of rapidity~\cite{Amelino-Camelia:2025ask}:
\begin{equation}
    U(\eta;M)=\frac{\sinh\eta [\cosh\eta\sinh{(\ell M)}+\cosh{(\ell M)}]}{\cosh\eta\cosh{(\ell M)}+\sinh{(\ell M)}}.\label{eq:velrapid}
\end{equation}
Note that this expression does not apply to massless particles, but it is convenient for describing the speed of the box.

From \myeq{eq:mass-shell} with \(m^2=0\) and \myeq{eq:speed}, expanding for \(\epsilon,\, p\ll 1/\ell\) yields the Taylor expansion of the speed of light:
\begin{equation}
    u(\epsilon)\simeq 1+\ell\epsilon+\mathcal{O}(\ell^2). 
\end{equation}
Thus, \(\ell<0\) corresponds to the subluminal case, while \(\ell>0\) corresponds to the superluminal case.
With this brief review, we now proceed to analyze the tension described in \mysec{sec:model-indep}.


\section{Quantitative Analysis in DSR1\label{sec:analysis}}
In this section, we analyze the speed of light in different inertial frames within the DSR1 framework.
We show that, within DSR1 and under the assumptions specified in \mysec{sec:model-indep}, a subluminal speed of light leads to a nontrivial kinematical tension of the type discussed therein.
For the superluminal case, a similar tension can also arise, but may be avoided under certain physically motivated assumptions.
The discussion of relative locality and generalizations to other models is deferred to \mysec{sec:discussion}.

\subsection{The subluminal case}
For the subluminal case, we take $\ell<0$.
Consider a box at rest in the frame $\oprime$, as depicted in Figs.~\ref{fig:fig1} and~\ref{fig:fig4}.
At $t'=0$, the sources $S_F$ and $S_B$ emit photons in the $+x'$ and $-x'$ directions, respectively.
The backward-moving photon merely ensures that the box remains at rest; its details are irrelevant to the analysis.

Let the energy and momentum of the forward-moving photon be $\epsilon'$ and $p'$.
We assume $\lvert\epsilon'\rvert, \lvert p'\rvert \lesssim 1/\lvert\ell\rvert$ throughout.
Since particle velocity and coordinate speed coincide in DSR1, the photon speed is $V(\epsilon',p') = \bar{x}^1/\bar{x}^0$.

Now consider an inertial observer in frame $\ocal$, in which the box moves with rapidity $\eta$ in the $+x$-direction, as shown in Figs.~\ref{fig:fig3} and~\ref{fig:fig5}.
Using \myeq{eq:coordtrans} and \myeq{eq:mass-shell}, the coordinate speed of the photon in $\ocal$ is
\footnote{The transformation from $\oprime$ to $\ocal$ is parameterized by $-\eta$, since the box has rapidity $\eta$ relative to $\ocal$.}
\begin{equation}
    V(\epsilon,p)=\frac{x^1(-\eta)}{x^0(-\eta)}=1+(e^{\ell\epsilon'}-1)e^{\eta},\label{eq:speedforocal}
\end{equation}
where the transformed energy $\epsilon$ and momentum $p$ are obtained from \myeq{eq:momentrans} with rapidity $-\eta$:
\begin{equation}
    \begin{aligned}
        \epsilon &=\epsilon'-\frac{1}{\ell}\ln\frac{e^{\ell\epsilon'}}{1-e^\eta+e^{\ell\epsilon'+\eta}}\simeq e^\eta\epsilon'+\mathcal{O}(\ell),\\
        p&=\frac{e^\eta-e^{\ell\epsilon'+\eta}}{\ell(1-e^\eta+e^{\ell\epsilon'+\eta})}\simeq -e^\eta\epsilon'+\mathcal{O}(\ell).\label{eq:epprime}
    \end{aligned}
\end{equation}
One readily verifies that these expressions satisfy \myeq{eq:speedforocal}.
We may therefore regard the photon speed as a function $V(\eta;\epsilon')$ of the rapidity.

The speed of the box in $\ocal$ follows from \myeq{eq:velrapid}:
\begin{equation}
    U(\eta;\massbox)=\sinh\eta\,
    \frac{\cosh\eta\sinh(\ell'\massbox)+\cosh(\ell'\massbox)}{\cosh\eta\cosh(\ell'\massbox)+\sinh(\ell'\massbox)},\label{eq:speedofbox}
\end{equation}
where $\massbox$ is the mass of the box.
The LSV scale $\ell'$ appearing in \myeq{eq:speedofbox} is \emph{not} necessarily equal to the microscopic scale $\ell$.
For a macroscopic object, the product $\ell\massbox$ can easily exceed unity (since $1/\lvert\ell\rvert$ is expected to be of order $E_{Pl}\sim 10^{-5}~\text{g}$), leading to macroscopic deviations from SR that are excluded by experiment.
This is the well-known \emph{soccer-ball problem} of DSR~\cite{Judes:2002bw,Magueijo:2002am,Magueijo:2006qd,Amelino-Camelia:2011dwc,Hossenfelder:2014ifa}, which originates from the curved geometry of momentum space.
Several proposals exist to address this issue, but no consensus has been reached.

Since experimental tests show no macroscopic deviation from SR, we first adopt the conservative assumption $\ell'=0$ in \myeq{eq:speedofbox}.
We subsequently consider the case $\ell'\neq 0$ and show that the qualitative behavior remains unchanged.

\subsubsection{Case \texorpdfstring{$\ell'=0$}{}}
Setting $\ell'\massbox=0$, the box speed reduces to
\[ U(\eta;\massbox)=\tanh\eta. \]
The question is whether there exists a physically attainable rapidity $\eta$ such that $V(\epsilon,p) < U(\eta;\massbox)$.

The critical rapidity $\eta_c$ at which $V=U$ is obtained by solving $1+(e^{\ell\epsilon'}-1)e^{\eta} = \tanh\eta$.
The solution is
\begin{widetext}
\begin{equation}
\begin{aligned}
\eta_c(\epsilon')=\ln \left(\frac{\left(\left(e^{\ell \epsilon' }-1\right)^2 \left(\sqrt{3} \sqrt{-2 e^{\ell \epsilon' }+e^{2 \ell
   \epsilon' }+28}-9\right)\right)^{2/3}+2 \sqrt[3]{3} e^{\ell \epsilon' }-\sqrt[3]{3} e^{2 \ell \epsilon'
   }-\sqrt[3]{3}}{3^{2/3} \left(e^{\ell \epsilon' }-1\right) \sqrt[3]{\left(e^{\ell \epsilon' }-1\right)^2
   \left(\sqrt{3} \sqrt{-2 e^{\ell \epsilon' }+e^{2 \ell \epsilon' }+28}-9\right)}}\right),\quad \ell<0.\label{eq:rapidsol}
\end{aligned}
\end{equation}
\end{widetext}

Two points must be verified: (i) that $U>V$ for $\eta>\eta_c$, and (ii) that $\eta_c$ lies within the allowed range of rapidity (since $\ell<0$ implies an upper bound on $\eta$).
Point (i) ensures that the speed of the box can actually exceed that of the photon, and point (ii) guarantees the mathematical consistency of the result.

\begin{figure}
    \centering
    \includegraphics[width=\linewidth]{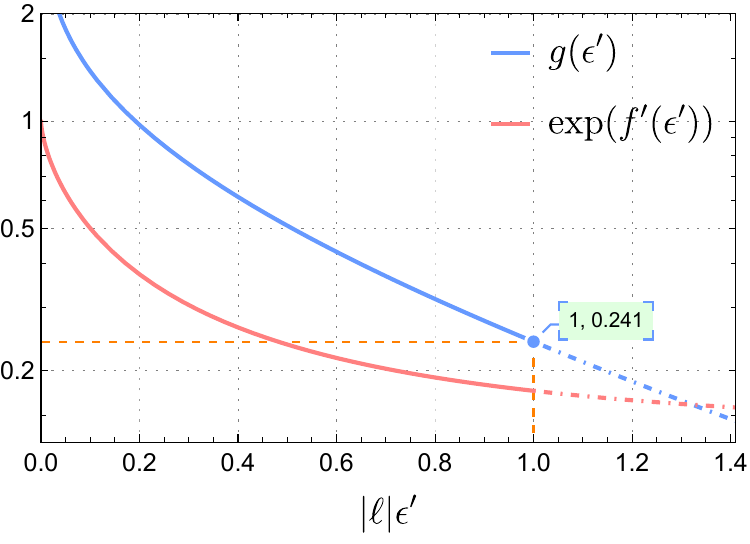}
    \caption{The functions $g(\epsilon') = \eta_+(\epsilon') - \eta_c(\epsilon')$ (solid) and $\exp[f'(\epsilon')]$ (dashed).
    The positivity of $g$ shows that the critical rapidity is attainable; the fact that $\exp(f') < 1$ shows that $f(\epsilon')$ is monotonically decreasing.}
    \label{fig:fig6}
\end{figure}

For point (i), define $f(\epsilon') = V(\eta;\epsilon') - U(\eta;\massbox)$ and differentiate with respect to $\eta$.
Figure~\ref{fig:fig6} shows $\exp[f'(\epsilon')]$ evaluated at $\eta=\eta_c$.
Since $\exp[f'(0)]=1$ and $f'(\epsilon') < 0$ for all $\epsilon'>0$, $f(\epsilon')$ is monotonically decreasing on $(0,+\infty)$.
Thus $V < U$ whenever $\eta > \eta_c$.

For point (ii), the transformation laws \myeq{eq:momentrans} and \myeq{eq:coordtrans} with $\ell<0$ imply that the rapidity cannot be arbitrarily large~\cite{Bruno:2001mw,Amelino-Camelia:2025ask}.
The bound is determined by the condition that the boosted energy and momentum remain finite, which gives
\begin{equation}
    \eta_\pm(\epsilon') = \mp\ln(1-e^{\ell\epsilon'}).
\end{equation}
The relevant upper bound is $\eta_+(\epsilon') = -\ln(1-e^{\ell\epsilon'})$~\footnote{This bound can be obtained directly by solving $p_0(-\eta)=+\infty$ or $p_1(-\eta)=-\infty$ in \myeq{eq:epprime}.}.

Defining $g(\epsilon') = \eta_+(\epsilon') - \eta_c(\epsilon')$, Fig.~\ref{fig:fig6} shows that $g(\epsilon') > 0$ for $0 < \epsilon' < 1/\lvert\ell\rvert$.
Hence $\eta_c$ lies within the physically accessible rapidity range.

\begin{figure}
    \centering
    \includegraphics[width=\linewidth]{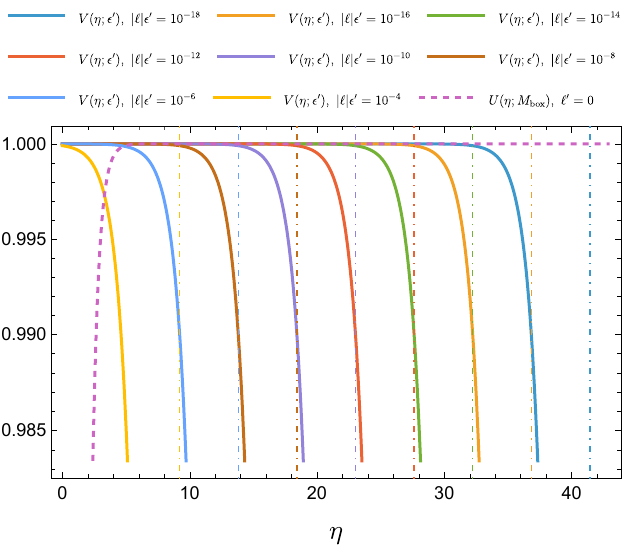}
    \caption{Comparison of the photon speed $V(\eta;\epsilon')$ (solid curves) and the box speed $U(\eta;\massbox)$ (dashed curve) for $\ell'=0$.
    Different colors correspond to different values of $\lvert\ell\rvert\epsilon'$ as indicated in the legend.
    The vertical dot-dashed lines mark the upper bounds $\eta_+(\epsilon')$ for each choice of $\epsilon'$ (same color coding).}
    \label{fig:fig7}
\end{figure}

For illustration, we take the LSV scale $1/\lvert\ell\rvert \sim 10^{17}~\text{GeV}$ (as suggested by subluminal linear-correction phenomenology), and choose representative photon energies $\epsilon' = 10^{-1}$, $10^1$, $10^3$, $10^5$, $10^7$, $10^9$, $10^{11}$, $10^{13}~\text{GeV}$.
Figure~\ref{fig:fig7} compares $V(\eta;\epsilon')$ (solid curves) with $U(\eta;\massbox)$ (dashed curve).
The vertical dot-dashed lines mark the upper bounds $\eta_+(\epsilon')$ for each energy.
In all cases, there exists a range $\eta_c < \eta < \eta_+$ where the box overtakes the photon, indicating the emergence of the kinematical tension described in \mysec{sec:model-indep}.

\subsubsection{Case \texorpdfstring{$\ell'\neq 0$}{}}
We now allow $\ell'\neq 0$, i.e., the macroscopic box obeys DSR1 kinematics with an effective scale $\ell'$.
The analysis is slightly more involved, but the qualitative conclusion remains unchanged.

For a macroscopic object, the combination $\ell'\massbox$ must be small; otherwise, macroscopic deviations from SR would have already been observed.
To estimate $\ell'\massbox$, suppose the box is composed of $N$ elementary particles, each of mass $m$ and each obeying DSR1 with the fundamental scale $\ell$.
An exemplary proposal for resolving the soccer-ball problem is that the effective scale for a composite object is $\ell' = \ell/N$.
Then $\ell'\massbox = \ell \massbox/N \le \ell m$, since the binding energy implies $\massbox \le N m$.

\begin{figure}
    \centering
    \includegraphics[width=\linewidth]{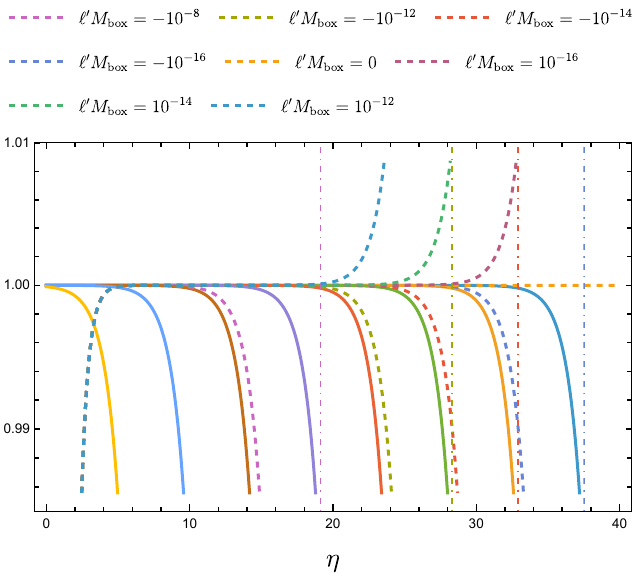}
    \caption{Same as Fig.~\ref{fig:fig7}, but with $\ell'\neq 0$.
    The solid curves are the photon speeds (identical to Fig.~\ref{fig:fig7}).
    The dashed curves show the box speed $U(\eta;\massbox)$ for different values of $\ell'\massbox$ (see legend).
    Vertical dot-dashed lines mark the rapidity bounds $\eta_{+,\text{box}}$ from \myeq{eq:rapidboundbox}.
    The bounds $\eta_+(\epsilon')$ from the photon sector are omitted for clarity, but can be read from Fig.~\ref{fig:fig7}.}
    \label{fig:fig8}
\end{figure}

Expanding \myeq{eq:speedofbox} for small $\ell'\massbox$ gives
\begin{equation}
    U(\eta;\massbox) \simeq \tanh\eta + \ell'\massbox \sinh\eta \tanh^2\eta + \mathcal{O}\left(\ell'^2\right).
\end{equation}
The correction is manifestly small.
Figure~\ref{fig:fig8} illustrates this: the solid curves (photon speeds) are identical to those in Fig.~\ref{fig:fig7}, while the dashed curves show $U(\eta;\massbox)$ for several values of $\ell'\massbox$ (including superluminal cases, shown only for completeness).

For the subluminal box, the rapidity is again bounded:
\begin{equation}
    \eta_{+,\text{box}} = \ln\frac{1+e^{\ell'\massbox}}{1-e^{\ell'\massbox}}.\label{eq:rapidboundbox}
\end{equation}
The vertical dot-dashed lines in Fig.~\ref{fig:fig8} indicate these bounds.
Even for $\lvert\ell'\rvert\massbox \sim 10^{-8}$, a sufficiently energetic forward photon still allows the box to overtake it.
A conservative estimate gives $\lvert\ell'\rvert\massbox \lesssim 10^{-14}$ (taking the new-physics scale $\sim 10^{16}~\text{GeV}$ and the electroweak scale $\sim 10^2~\text{GeV}$), which lies well within the regime where the above behavior persists.

Thus, the effect is robust: even with a nonzero effective DSR scale for macroscopic objects, the boosted box can overtake its own photon in $\ocal$ within the present framework.

\subsection{The superluminal scenario}
One might expect that if photons are superluminal in DSR1 ($\ell>0$), the tension disappears.
This is not automatically true, but the issue can be avoided under certain physically reasonable conditions.

The photon speed is again given by \myeq{eq:speedforocal}:
\begin{equation}
    V(\eta;\epsilon') = 1 - e^\eta + e^{\ell\epsilon'+\eta},
\end{equation}
now with $\ell\epsilon' > 0$.

For brevity, set $x = \ell\epsilon'$, $y = \eta$, and $z = \ell'\massbox$.
If macroscopic objects obey standard SR ($\ell'=0$), then $U(\eta;\massbox) = \tanh y$.
Since
\begin{equation}
    1 + (e^x - 1)e^y > 1 > \tanh y \qquad (x>0),\label{eq:ineq}
\end{equation}
we have $V > U$ for all $y$, and no tension arises.

If instead macroscopic objects follow DSR1 with $\ell' < 0$, we have the inequality
\begin{equation}
    \frac{\cosh z + \sinh z \cosh y}{\cosh z + \frac{\sinh z}{\cosh y}} < 1 \qquad (z<0,\ 0<y<\ln\tfrac{1+e^z}{1-e^z}),\label{eq:ineq2}
\end{equation}
which, together with \myeq{eq:speedofbox}, implies $U(\eta;\massbox) < \tanh\eta$.
Combined with \myeq{eq:ineq}, this again guarantees $V > U$ over the allowed parameter range.
Note that the rapidity bound in \myeq{eq:ineq2} is precisely the upper limit \myeq{eq:rapidboundbox} for the box, so the inequality holds over the entire physical range.

The problematic case is $\ell'>0$.
Here the inequality reverses:
\begin{equation}
    \frac{\cosh z + \sinh z \cosh y}{\cosh z + \frac{\sinh z}{\cosh y}} > 1 \qquad (z>0,\ y>0).\label{eq:ineq3}
\end{equation}
For sufficiently soft photons ($x\sim 0$), the box can overtake the photon even though both are superluminal.
A numerical example illustrates this: with $\ell\epsilon' = 10^{-18}$ and $\ell'\massbox = 10^{-16}$, $U > V$ for $\eta \gtrsim 13$.

Nevertheless, this issue can be remedied.
One simple option, for example, is to require $\ell' \le 0$ for macroscopic objects; the above analysis then guarantees $V > U$ for all photon energies.
A more plausible resolution is to assume a lower bound on physical photon energies---either a universal infrared cutoff or a nonzero photon rest mass~(for example, see Ref.~\cite{Tu:2005ge,Goldhaber:2008xy}).
In this case, $\ell\epsilon'$ has a positive minimum $x_{\min}$, and we expect $\ell'\massbox \le x_{\min}$ for macroscopic objects.

To verify that this prevents the inconsistency, define
\begin{equation}
    h(x,y,z) = 1 - e^y + e^{x+y} - \sinh y \, \frac{\sinh z \cosh y + \cosh z}{\cosh y \cosh z + \sinh z},\label{eq:funch}
\end{equation}
with $x, y, z > 0$ and $x > z$ (the latter reflecting the assumed infrared cutoff).
Since $\partial_x h(x,y,z) > 0$, $h$ is monotonically increasing in $x$, and its infimum is $h(z,y,z)$.
As shown in Appendix~\ref{sec:app}, $h(z,y,z) > 0$ for all $y,z > 0$.
Hence $h(x,y,z) > 0$, implying $V > U$ for all allowed parameters.

In summary, the superluminal scenario of DSR1 does not automatically eliminate the tension.
However, unlike the subluminal case, it admits physically plausible conditions (e.g., $\ell' \le 0$ or an infrared cutoff) under which the tension is avoidable.


\section{Discussion and Summary\label{sec:discussion}}

\subsection{Relative locality does not resolve the inconsistency}
The paradoxes identified in Refs.~\cite{Hossenfelder:2009mu,Hossenfelder:2010tm} can be understood, at least in part, as manifestations of relative locality~\cite{Amelino-Camelia:2010wxx,Amelino-Camelia:2011lvm,Amelino-Camelia:2011ebd,Amelino-Camelia:2011dwc,Amelino-Camelia:2011hjg}---the idea that the curved geometry of momentum space implies that different inertial observers may disagree on the localization of events.
It is therefore natural to ask whether the tension found in this work can be resolved in a similar way.
We now argue that this is not the case.

Relative locality allows events that are coincident in one frame to appear spatially separated in another.
In particular, the emission of the forward-moving photon, which occurs at the box wall in $\oprime$, may be assigned a shifted position in $\ocal$, either in front of or behind the wall.
At the level of individual events, this seems to restore consistency between observers.
However, the tension identified in this work does not hinge on the precise localization of the emission event, but rather on the global structure of particle propagation.
The tension is therefore insensitive to the assignment of emission coordinates; it resides instead in the global causal structure of worldlines and their intersections.

Once the box overtakes the forward photon in $\ocal$, inconsistencies arise at the level of asymptotic observables, independently of where the emission event is localized.
To make this explicit, consider the two possible assignments:

\begin{itemize}
    \item \emph{The photon is emitted in front of the wall.}
    In $\ocal$, the photon is created at time $t_0$ ahead of the box.
    Since the box moves faster than the photon, it will eventually catch up and intersect its worldline at some later time $t_1$.
    The observational history in $\ocal$ is then discontinuous: no photon before $t_0$, two photons between $t_0$ and $t_1$, and only one photon after $t_1$.
    In contrast, in $\oprime$ both photons propagate indefinitely after emission.
    This leads to a direct disagreement in the asymptotic particle number.

    \item \emph{The photon is emitted behind the wall.}
    In this case, the photon is created behind the box in $\ocal$ and never catches up with it.
    One might then conclude that two photons persist indefinitely.
    However, if one introduces a detector in front of the box that absorbs the forward photon, the inconsistency reappears.
    In $\oprime$, the photon is emitted and later absorbed, leaving only the backward photon asymptotically.
    In $\ocal$, the photon never reaches the detector and is never absorbed.
    Again, the two observers disagree on the final particle content.
\end{itemize}

These inconsistencies arise at the level of asymptotic detection and cannot be eliminated by shifting the localization of events.
Thus, relative locality does not resolve the inconsistency; it merely relocates it from event coincidence to particle interpretation.
Unlike the situations discussed in Refs.~\cite{Hossenfelder:2009mu,Hossenfelder:2010tm}, the tension here is not tied to local coincidences, but to the global consistency of particle propagation and counting.
This indicates a more severe conflict between observer-independent LSV and relativistic covariance.

\subsection{Generalization to more general models}
Although our quantitative analysis was performed within the DSR1 model, the origin of the tension is not specific to this realization.
The essential ingredients are:
(i) an observer-independent functional relation between speed and energy, $u(E)$, and
(ii) a deformed boost transformation that preserves this functional form across inertial frames.

We stress that monotonicity of $u(E)$ is not imposed as an independent assumption.
Rather, in all presently known DSR constructions with light-speed variation, the photon energy transforms monotonically with rapidity within the regime of validity of the effective description.
Under this condition, the covariant relation $u(E)$ necessarily induces a corresponding systematic variation of the photon speed under boosts.

Therefore, the effect identified here does not rely on a special choice of model, but follows from the combination of observer-independent LSV and monotonic energy growth under boosts---features common to existing DSR realizations with light-speed variation.
Once both conditions are imposed, the speed of a photon in a boosted frame is entirely determined by its transformed energy and cannot be adjusted independently.
As a result, if the speed decreases with energy, a sufficiently large boost inevitably leads to a regime in which a macroscopic object overtakes its own radiation.

The tension can therefore be traced to the simultaneous imposition of covariance and energy-dependent propagation.
Since covariance is a defining principle of DSR, this tension reflects a structural incompatibility rather than a peculiarity of a specific realization.
This does not imply that all DSR models are ruled out.
For example, in DSR2 the photon dispersion relation~\cite{Magueijo:2001cr,Magueijo:2002am}
\begin{equation}
    \frac{E^2}{(1-E/\elsv)^2} - \frac{p^2}{(1-E/\elsv)^2} = 0,
\end{equation}
implies an energy-independent speed of light, and no tension arises.

More generally, models with superluminal LSV can remain consistent provided additional conditions are imposed, as discussed in \mysec{sec:analysis}.
Indeed, a superluminal LSV in DSR is somewhat preferred on theoretical grounds, mainly because the rapidity remains unbounded in this case~\cite{Amelino-Camelia:2003xax}.
The obstruction identified here is therefore specific to subluminal, covariant LSV.

\subsection{A signal of \textit{doubly general relativity}?}
Another possible interpretation is that the inconsistency signals the breakdown of the DSR framework at large boosts.
DSR is a kinematic modification of special relativity, formulated in flat spacetime but with a nontrivial momentum-space geometry.
As such, it is expected to be an effective description valid only within a certain regime.

In the scenario considered here, the inconsistency arises when the rapidity exceeds the critical value $\eta_c$.
In this regime, the boosted energies can become large, potentially probing the domain where a full dynamical theory of quantum spacetime is required.
From this perspective, the tension may indicate that a purely kinematic deformation is insufficient, and that spacetime and momentum space must be treated on equal dynamical footing.
This suggests the need for a more complete framework, sometimes referred to as \emph{doubly general relativity} (DGR), in which both spacetime curvature and momentum-space curvature are incorporated consistently.

While no complete formulation currently exists, the present inconsistency provides a concrete diagnostic of where the DSR description may fail.

\subsection{Summary and outlook}
In this work, we have analyzed the consistency of observer-independent light-speed variation (LSV) within doubly special relativity (DSR).
Using a simple thought experiment involving an inertial box emitting two photons, we have shown that subluminal LSV leads to a robust and quantitative tension.
The key result is that, in a sufficiently boosted frame, the box can overtake its own forward-emitted photon.
This leads to inconsistencies in asymptotic particle counting and in the identification of inertial motion.

We have demonstrated explicitly, within the DSR1 model, that the critical rapidity $\eta_c$ at which this occurs lies within the physically allowed range of deformed boosts, and that the effect persists even when macroscopic corrections are included.
We have further shown that this tension cannot be resolved by relative locality.
The inconsistency survives at the level of asymptotic observables and reflects a tension between three ingredients: observer-independent LSV, relativistic covariance, and standard notions of particle propagation.

While superluminal scenarios can evade the inconsistency under additional assumptions, subluminal covariant LSV appears to be intrinsically problematic.
This places nontrivial constraints on quantum-gravity phenomenology based on such effects.

Several directions merit further investigation.
It would be important to test the generality of this result in other DSR models, and to clarify the role of multi-particle states and the effective scale $\ell'$ in macroscopic systems.
More broadly, the inconsistency may serve as a guide for developing frameworks beyond DSR, in which the interplay between spacetime structure and momentum-space geometry is treated dynamically.


\section*{Acknowledgments}

The authors would like to acknowledge the valuable discussions with Bo-Qiang Ma.
This work was supported in part by the National Natural Science Foundation of China~(NSFC) under Grant No.~12547101. HL was also supported by the start-up fund of Chongqing University under No.~0233005203009, and JZ was supported by the start-up fund of Chongqing University under No.~0233005203006.


\appendix

\section{Proof of \texorpdfstring{\(h(z,y,z)>0\)}{} for \texorpdfstring{\(y,z>0\)}{}\label{sec:app}}

In the following, \(y,z>0\) is understood.
According to \myeq{eq:funch}, the denominator of \(h(z,y,z)\) is
\begin{equation}
    h_d(y,z):=8 e^{2 y + z} (\cosh (y) \cosh (z) + \sinh (z)),\label{eq:appendixeq1}
\end{equation}
which is already positive. The numerator is
\begin{widetext}
    \begin{equation}
        \begin{aligned}
            h_n(y,z)&:=2 \left(e^y-1\right)^2 e^{2 y+z}-\left(e^y+1\right)^2 \left(3 e^y+1\right) \left(e^y-1\right) e^{2 z}+2 \left(e^y+1\right)^2 e^{2 y+3 z}-\left(e^y-1\right)^4\\
            =&\underbrace{e^{4 y} \left(2 e^{3z}-3 e^{2 z}+2 e^{z}-1\right)}_{\ge 0\text{ for }z\ge 0}+\underbrace{4 e^{3 y} \left(e^z+1\right) \left(e^z-1\right)^2}_{>=0}+\underbrace{4 e^y \left(e^{2 z}+1\right)}_{>0}+\underbrace{2 e^{2 y} \left(e^{3z}+e^{2 z}+e^{z}-3\right)}_{\ge 0\text{ for }z\ge0}+\underbrace{\left(e^{2 z}-1\right)}_{\ge 0\text{ for }z\ge 0},
            \label{eq:appendixeq2}
        \end{aligned}
    \end{equation}
\end{widetext}
which ensures that \(h_n(y,z)>0\) for \(y,z>0\).
Since we have \(h(z,y,z)=h_n(y,z)/h_d(y,z)\), this completes our proof of \(h(z,y,z)>0\) for \(y,z>0\).


\bibliography{refs}

\end{document}